\documentclass{article}
\usepackage{spconf,amsmath,graphicx}
\usepackage{multirow}
\usepackage{caption}
\usepackage{cite}
\usepackage{url}
\usepackage{enumitem}
\usepackage{adjustbox}

\title{LIBRIHEAVY: A 50,000 HOURS ASR CORPUS WITH PUNCTUATION CASING AND CONTEXT}
\name{Wei Kang, Xiaoyu Yang, Zengwei Yao, Fangjun Kuang, Yifan Yang, Liyong Guo, Long Lin, Daniel Povey}

\address{Xiaomi Corp., Beijing, China \\
\footnotesize{\texttt{\{kangwei1, dpovey\}@xiaomi.com}} \vspace{-1em}}

\begin{document}
%
\maketitle
\begin{abstract}

In this paper, we introduce Libriheavy, a large-scale ASR corpus consisting of 50,000 hours of read English speech derived from LibriVox. To the best of our knowledge, Libriheavy is the largest freely-available corpus of speech with supervisions. Different from other open-sourced datasets that only provide normalized transcriptions, Libriheavy contains richer information such as punctuation, casing and text context, which brings more flexibility for system building. Specifically, we propose a general and efficient pipeline to locate, align and segment the audios in previously published Librilight to its corresponding texts. The same as Librilight, Libriheavy also has three training subsets small, medium, large of the sizes 500h, 5000h, 50000h respectively. We also extract the dev and test evaluation sets from the aligned audios and guarantee there is no overlapping speakers and books in training sets. Baseline systems are built on the popular CTC-Attention and transducer models. Additionally, we open-source our dataset creatation pipeline which can also be used to other audio alignment tasks.

\end{abstract}

\vspace{-4mm}
\section{Introduction}
\vspace{-2mm}

In the past decade, various system architectures, like Connectionist Temporal Classification (CTC)~\cite{ctc}, RNN-T~\cite{transducer} and encoder-decoder based model~\cite{chan2015listen}, have been proposed, pushing the dominant framework from the hybrid Hidden Markov Models (HMM)~\cite{dahl2011context} to end-to-end models. In general, the neural network models are supposed to be more data hungry than traditional systems.

A lot of work has been done on publishing open source datasets, for example, the Wall Street Journal corpus~\cite{wall-street-journal}, SwitchBoard~\cite{switchboard}, Fisher~\cite{fisher} and the famous LibriSpeech corpus~\cite{librispeech}. While these are all small or medium size datasets with less than 2,000 hours, which is too small to train a good enough end-to-end model. In recent years, there are also large-scale corpora like GigaSpeech~\cite{gigaspeech}, People's Speech~\cite{peoplespeech} and MLS~\cite{mlspeech}. One drawback of these datasets is that they only provide normalized transcriptions, making it impossible to train a model that needs full-format texts, such as punctuation prediction.  

Typical ASR corpora aim at training ASR systems to recognize independent utterances. However, the preceding context of the current utterance may convey useful information. Contextualized speech recognition utilizes the cross-utterance context to improve the accuracy of ASR systems and yields promising results~\cite{wei2021attentive, SelfAttnBiasing}. However, training such systems usually requires utterance-level context for each training utterance, which is not available in most existing ASR corpora. Therefore, such a dataset with textual context information is highly desirable.


Motivated by the aforementioned  points, we introduce Libriheavy, a large-scale (50,000 hours) corpus containing not only fully formatted transcripts but also textual context, which is suitable for various speech recognition related tasks. In addition, unlike other open-source datasets that have their own creating pipelines, we propose a general audio alignment method and release it as a standard package. Our contributions are as follows:
\begin{itemize}
[leftmargin=*,topsep=2pt,parsep=0pt,itemsep=2pt,]
    \item We release a 50,000-hour of labeled audio containing punctuation casing and preceding text;
    \item We propose and open-source a general audio alignment pipeline, which makes it easier to construct ASR corpora;
    \item We provide solid evaluation results on Libriheavy, which demonstrate the high quality of the corpus and the robustness of our pipeline.
\end{itemize}


\vspace{-4mm}
\section{Libriheavy corpus}
\vspace{-2mm}
In this section, we provide a detailed description of the Libriheavy corpus, including audio files, metadata, data partitions, text styles, and other aspects. Instructions and scripts are available in the Libriheavy GitHub repository\footnote{\url{https://github.com/k2-fsa/libriheavy}}.

\vspace{-4mm}
\subsection{Librilight}
\vspace{-2mm}

Librilight~\cite{librilight} is a collection of unlabeled spoken English audio derived from open-source audio books from the LibriVox project \footnote{\url{https://librivox.org}}. It contains over 60,000 hours of audio and aims for training speech recognition systems under limited or no supervision. The corpus is free and publicly available~\footnote{\url{https://github.com/facebookresearch/libri-light}}.


\vspace{-4mm}
\subsection{Libriheavy}
\vspace{-2mm}

Libriheavy is a labeled version of Librilight. We align the audio files in Librilight to their corresponding text in the original book and segment them into smaller pieces with durations ranging from 2 to 30 seconds.
We maintain the original dataset splits of Librilight and have three training subsets (small, medium, large). In addition, we further extract evaluation subsets (dev, test-clean, test-other) for validation and testing. Table~\ref{tab:statistic} shows the statitics of these subsets.

\begin{table}[t]
\centering
\caption{The dataset statistics of Libriheavy.}
\vspace{-2mm}
\label{tab:statistic}
\begin{tabular}{lllll}
\hline
subset                  & hours                  & books                  & per-spk hrs                  & total spks     \\ \hline
small                   &   509                  &   173                  &   1.22                         &  417         \\
medium                  &   5042                 &   960                  &   3.29                         &  1531        \\
large                   &   50794                &   8592                 &   7.54                         &  6736        \\ \hline
dev                     &   22.3                 &   180                  &   0.16                         &  141         \\
test-clean              &   10.5                 &   87                   &   0.15                         &  70          \\
test-other              &   11.5                 &   112                  &   0.16                         &  72          \\ \hline
\end{tabular}
\vspace{-4mm}
\end{table}

\vspace{-4mm}
\subsubsection{Metadata}
\vspace{-2mm}

We save the metadata of the dataset as Lhotse~\cite{lhotse} cuts in JSON lines. Each line is a self-contained segment, including the transcript and its audio source. Users can clip the corresponding audio segment with the given \textit{start} and \textit{duration} attributes.
Unlike other publicly available corpora that only provide normalized transcripts, Libriheavy includes richer information such as punctuation, casing, and text context.
The text context is the transcription of the preceding utterances,
located in the \textit{pre\_texts} entry, with a default length of 1000 bytes. There are also \textit{begin\_byte} and \textit{end\_byte} attributes, which allow users to easily slice any length of text context from the original book pointed to by the \textit{text\_path} attribute.
Of course, there are other supplementary entries that might be usefull for other tasks, such as \textit{id}, \textit{speaker}, etc.

%

\vspace{-4mm}
\subsubsection{Evaluation Sets}
\vspace{-2mm}

As mentioned above, we have three evaluation sets in Libriheavy, namely dev, test-clean, test-other. We ensure that the evaluation sets have no overlapping speakers and books in the training set. To make the evaluation sets contain as many speakers and books as possible while not dropping out too much training data, we filtered out speakers and books with shorter durations as candidates. We then determine the \textit{clean} speakers and \textit{other} speakers using the same method as in ~\cite{librispeech} and divide the candidates into \textit{clean} and \textit{other} pool.
We randomly select 20 hours of audio from the \textit{clean} pool, half of which forms the test-clean set and the other half is appended to the dev set. We follow the same procedure for the \textit{other} pool. Librilight ensures that audio files from the LibriSpeech evaluation sets are not present in the corpus, therefore, the LibriSpeech evaluation sets can also be used as our evaluation sets.

\vspace{-4mm}
\section{Audio Alignment}
\vspace{-2mm}

This section describes the creation pipeline of the Libriheavy corpus. The key task of audio alignment is to align the audio files to the corresponding text and split them into short segments, while also excluding segments of audio that do not correspond exactly with the aligned text. Our solution presented here is a general pipeline that can be applied to other data generation tasks as well. The implementation of all the following algorithms and corresponding scripts are publicly available\footnote{\url{https://github.com/k2-fsa/text_search}}.

\vspace{-4mm}
\subsection{Downloading text}
\vspace{-2mm}

To align the audio derived from audiobooks, we require the original text from which the speaker read the audiobook. From the metadata provided by Librilight, we can obtain the URL of the textbook for each audio file. We have written scripts to automatically extract the text and download the sources for all audiobooks. We then apply simple clean-up procedures such as removing redundant spaces and lines to the text sources.

\vspace{-4mm}
\subsection{First alignment stage}

The goal of this stage is to locate the audio to its corresponding text segments (e.g. chapter) in the original book. First, we obtain the automatic transcript of the audio file. Then we treat the automatic transcript as \textbf{query} and the text in the original book as \textbf{target}~\footnote{We will normalize the text to upper case and remove the punctuation, but keep the index into original text.}, and find the close matches (Sec~\ref{ch:close_matches}) for each elements in the query over the target. Finally, we determine the text segment of the audio by finding the longest increasing pairs (Sec~\ref{ch:lingest_pairs}) of query elements and their close matches. Note, we did not use the VAD tool provided by Librilight for audio segmenting and as our algorithm requires a relatively long text to guarantee its accuracy.

\vspace{-4mm}
\subsubsection{Transcribe audios}
\vspace{-2mm}

The audios in Librilight have a large variance in duration, from a few minutes to hours. To avoid excessive computation on long audio files, we first split the long audio into 30-second segments with 2 seconds of overlap at each side, and then recognize these segments with an ASR model trained on Librispeech. Finally, we combine the transcripts that belong to the same audio by leveraging the timestamps of the recognized words.


\vspace{-4mm}
\subsubsection{Close matches}
\label{ch:close_matches}
\vspace{-2mm}

\begin{figure}
    \centering
    \includegraphics[width=0.45\textwidth]{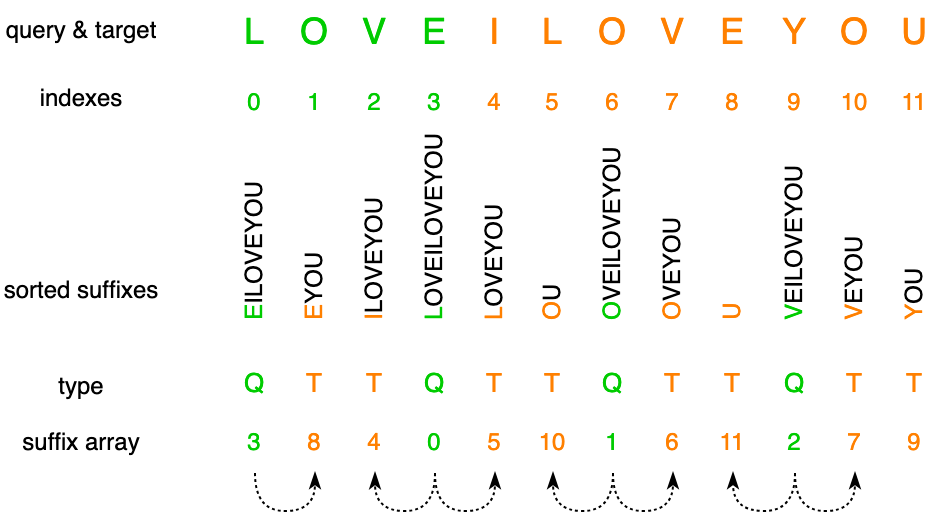}
    \caption{Example of finding close matches for a query (\textit{LOVE}) over the target (\textit{ILOVEYOU}). The dash arrows point from the query elements to their close matches.}
    \label{fig:suffix-array}
\vspace{-4mm}
\end{figure}

Now we have the automatic transcript and the original book for each audio. To obtain the most similar text segment in the original book of the automatic transcript roughly, we propose the \textit{close matches}. First, we concatenate query and target to a long sequence (target follows query), then a suffix array is constructed on the sequence using the algorithm in~\cite{2006linearsuffixarray}. The \textit{close matches} of the element in query position $i$ is defined as two positions in the original sequence that are within the target portion, and which immediately follow and precede, in the suffix array, query position $i$. This means that the suffixes ending at those positions are reverse-lexicographically close to the suffix ending at position $i$. Figure~\ref{fig:suffix-array} shows a simple example of finding the close matches of query \textit{``LOVE"} over target \textit{``ILOVEYOU"}.

\vspace{-4mm}
\subsubsection{Longest increasing pairs}
\label{ch:lingest_pairs}
\vspace{-2mm}

Let us think of those \textit{close matches} which we obtained above as a set of $\left(i, j\right)$ pairs, where $i$ is an index into query sequence and $j$ is an index into the target sequence. The query and its corresponding segment in the target should be monotonic aligned, so we can get the approximate alignment between the two sequences by finding the longest chain of pairs: $\left(i_1, j_1\right), \left(i_2, j_2\right), ... \left(i_N, j_N\right)$, such that $i_1 <= i_2 <= ... <= i_N$, and $j_1 <= j_2 <= ... <= j_N$. 




\vspace{-4mm}
\subsection{Second alignment stage}

From the longest chain obtained from the previous step, we can roughly locate the region in the target sequence relative to the query. At this stage, we use the Levenshtein alignment~\cite{levenshtein1966binary} to find the best single region of alignment between the recognized audio (query) and the text segment (obtained by the longest chain pairs). Since Levenshtein alignment is a quadratic time complexity algorithm, and will be very inefficient for long sequences. We can use the traceback through the pairs in the longest chain as the backbone for the Levenshtein alignment, so that we limit the Levenshtein alignment into blocks defined by the $\left(i, j\right)$ positions in this traceback. By concatenating the Levenshtein alignments of all the blocks along the query index, we obtain the Levenshtein alignment of the whole query.

\vspace{-4mm}
\subsection{Audio segmentation}
\vspace{-2mm}

The goal of audio segmentation is to break long audio into shorter segments, ranging from 2 seconds to 30 seconds, which are more suitable for ASR training. We use a two-stage scoring method to search for good segmentations~\footnote{The scores mentioned below will be normalized to the same scale, so none of the scores would dominate the final score.}. All books in LibriVox have punctuation, so we decided to split the sentence only at punctuation indicating the end of a sentence, namely, ``.”, ``?” and ``!” \footnote{Our toolkit also supports splitting sentences at a certain threshold of silence.}.  We select the positions in the alignment that follow chosen punctuations as Begin Of a Segment (BOS) and the positions followed by chosen punctuations as End Of a Segment (EOS), then we compute scores for these positions:
\begin{itemize}
[leftmargin=*,topsep=1.8pt,parsep=0pt,itemsep=1.8pt,]
    \item The number of silence seconds this position follows or is followed by, up to 3 seconds.
    \item The score corresponding to the number of insertions, deletions and substitutions within a certain region of this position.
\end{itemize}

Each pair of BOS and EOS forms a segment. The following rule is applied to assign scores to potential segments:
\begin{itemize}
[leftmargin=*,topsep=1.8pt,parsep=0pt,itemsep=1.8pt,]
    \item The score of BOS plus the score of EOS.
    \item A score related to the duration of the segment, which guarantees the duration is in the range of 2 to 30 seconds and encourages a duration between 5 to 20 seconds.
    \item A bonus for the number of matches in the alignment.
    \item A penalty for the number of errors in the alignment.
\end{itemize}

For each BOS, we find the 4 best-scoring EOS and vice versa. We then append the preceding 2 sets of segments to get a list of candidate segments. We determine the best segmentations by getting the highest-scoring set of segments that do not overlap. In practise, to avoid dropping out too much audio, we allow some kind of overlap if the overlapping length is less than a quarter of the segment.

\vspace{-2mm}
\section{Experiments}

In this section, we present the baseline systems and experimental results for two popular models, namely CTC-Attention~\cite{watanabeHybrid} and neural transducer~\cite{transducer}. We then compare the performance between the models trained on normalized text and texts with punctuation and casing. 

\vspace{-3mm}
\subsection{CTC-Attention baseline system}

We build the CTC-Attention baseline using the Wenet~\cite{wenet} framework.
We use the classic setup of Wenet toolkit which consists of a 12-layer Conformer~\cite{conformer} encoder and a 6-layer Transformer decoder. The embedding dimension is set to 512. The kernel size of the convolution layers is set to 31. The feedforward dimension is set to 2048. The modeling units are 500-class Byte Pair Encoding (BPE)~\cite{bpe} word pieces.
The loss function is a logarithmic linear combination of the CTC loss (weight = 0.3) and attention loss with label smoothing (weight = 0.1). The input features are 80-channel Fbank extracted on 25 ms windows shifted by 10 ms with dither equal 0.1.
SpecAugment~\cite{SpecAugment}
and on-the-fly Speed perturbation~\cite{speed_perturb} are also applied to augment the training data. During training, we use the Adam optimizer~\cite{2014Adam} with the maximum learning rate of 0.002. We use the Noam~\cite{vaswani2017transformer} learning rate scheduler with 25k warm-up steps.

The model is trained for 90, 60 and 15 epochs on the small, medium and large subsets, respectively.
Table~\ref{tab:ctc-system} shows the Word Error Rate (WER) of the models on Libriheavy test sets. As a reference, we also show the WER on the LibriSpeech test sets. The N-Best hypotheses are first generated by the CTC branch and then rescored by the attention branch. Note that for the LibriSpeech results, we apply some simple text normalization, such as converting numbers to their corresponding text and converting abbreviations (e.g ``Mr." to ``Mister") on the hypotheses to make it compatible with the LibriSpeech transcripts. We also apply these normalization procedures in the following experiments.

\begin{table}[]
\centering
\caption{The WERs of LibriSpeech (ls) and Libriheavy (lh) test sets on CTC-Attention system.}
\vspace{-2mm}
\label{tab:ctc-system}
\begin{tabular}{lllll}
\hline
subset              & ls-clean & ls-other & lh-clean & lh-other \\ \hline
small               &  5.76    &  15.60   &  6.94    & 15.17    \\ 
medium              &  3.15    &  7.88    &  3.80    & 8.80     \\ 
large               &  2.02    &  5.22    &  2.74    & 6.68     \\ \hline
\end{tabular}
\vspace{-2mm}
\end{table}


\vspace{-2mm}
\subsection{Transducer baseline system}

We build the transducer baseline system using the icefall~\footnote{\url{https://github.com/k2-fsa/icefall}} which is one of the projects in the Next-gen Kaldi toolkit. Icefall implements a transformer-like transducer system, which consists of a encoder and a stateless decoder~\cite{rnntstateless}. Different from the setting in ~\cite{rnntstateless} which only has an embedding layer, an extra 1-D convolution layer with a kernel size of 2 is added on top of it. The encoder used in this baseline is a newly proposed model called Zipformer~\cite{yao2023zipformer}. We use the default setting in the Zipformer LibriSpeech recipe~\footnote{\url{https://github.com/k2-fsa/icefall/blob/master/egs/librispeech/ASR/zipformer/zipformer.py}} in icefall for all the following experiments.

The same as CTC-Attention baseline system, we train the model for 90, 60 and 15 epochs for the small, medium and large subsets, respectively. Table~\ref{tab:transducer-system} shows the decoding results of the models trained on different training subsets, and the WERs on LibriSpeech and Libriheavy test sets are presented. We use the beam search method proposed in~\cite{kang2023fast} which limits the maximum symbol per frame to one to accelerate the decoding.

\begin{table}[t]
\centering
\caption{The WERs of LibriSpeech (ls) and Libriheavy (lh) test sets on transducer system.}
\vspace{-2mm}
\label{tab:transducer-system}
\begin{tabular}{lllll}
\hline
subset             & ls-clean & ls-other & lh-clean & lh-other \\ \hline
small              &  4.05    &  9.89    &  4.68    & 10.01    \\
medium             &  2.35    &  4.82    &  2.90    & 6.57     \\
large              &  1.62    &  3.36    &  2.20    & 5.57     \\ \hline
\end{tabular}
\vspace{-2mm}
\end{table}


\vspace{-2mm}
\subsection{Training with punctuation and casing}

This section benchmarks the performance of models trained on texts with punctuation and casing, and compares them with the performance of models trained on normalized texts. The system setting
is almost the same as the transducer baseline system mentioned above. The only difference is that we adopt 756-class BPE word pieces rather than 500 for modeling, because we open the \textit{fallback\_bytes} flag when training the BPE model to handle rare characters, so we need an additional 256 positions for bytes. Table~\ref{tab:transducer-cases-punc} shows the WERs and Char Error Rate (CER) of models trained on texts with punctuation and casing.

Table~\ref{tab:comparsion-cases-punc} compares the results of systems trained on normalized texts (upper case without punctuation) and unnormalized texts (casing with punctuation). In this experiment, we normalized both the transcripts and decoding results to upper case and removed the punctuation when calculating the WERs. From the results, the performance gap between two types of training texts is large when the training set is small, but as the training set grows, the gap becomes negligible. This indicates that when the training set is large enough, the style of training texts will not make much difference on performance, while training with texts with punctuation and casing brings us more information and flexibilities.


\begin{table}[]
\centering
\caption{The Libriheavy WERs and CERs on transducer system trained on texts with punctuation and casing.}
\vspace{-2mm}
\label{tab:transducer-cases-punc}
\begin{tabular}{lllll}
\hline
\multirow{2}{*}{subset} & \multicolumn{2}{c}{WER} & \multicolumn{2}{c}{CER} \\ \cline{2-5} 
                        & lh-clean   & lh-other   & lh-clean   & lh-other   \\ \hline
small                   &  13.04     & 19.54      &  4.51      &  7.90      \\
medium                  &  9.84      & 13.39      &  3.02      &  5.10      \\
large                   &  7.76      & 11.32      &  2.41      &  4.22      \\ \hline
\end{tabular}
\vspace{-2mm}
\end{table}

\begin{table}[t]
\centering
\caption{The comparison of WERs between models trained on Upper case No Punctuation (UNP) and Casing with Punctuation (C\&P).}
\vspace{-2mm}
\label{tab:comparsion-cases-punc}
\adjustbox{max width=\linewidth}{
\begin{tabular}{llllll}
\hline
subset                   & text          & ls-clean & ls-other & lh-clean & lh-other \\ \hline
\multirow{2}{*}{small}   & UNP           &  4.05    &  9.89    &  4.68    &  10.01   \\ 
                         & C\&P          &  4.51    &  10.84   &  5.16    &  11.12   \\ \hline
\multirow{2}{*}{medium}  & UNP           &  2.35    &  4.82    &  2.90    &  6.57    \\  
                         & C\&P          &  2.45    &  5.03    &  3.05    &  6.78    \\ \hline
\multirow{2}{*}{large}   & UNP           &  1.62    &  3.36    &  2.20    &  5.57    \\ 
                         & C\&P          &  1.72    &  3.52    &  2.28    &  5.68    \\ \hline
\end{tabular}
}
\vspace{-4mm}
\end{table}

\vspace{-2mm}
\section{Conclusion}

We release a large-scale (50,000 hours) corpus containing punctuation, casing and text context, which can be used in various of ASR tasks. We also propose and open-source a general and efficient audio alignment toolkit, which makes constructing speech corpora much easier. Finally, we conduct solid experiments on the released corpus, and the results show that our corpus is of high quality and demonstrates the effectiveness of our creation pipeline.

\clearpage
\bibliographystyle{IEEEbib}
\bibliography{refs}

\end{document}